\documentclass[twocolumn,aps,prc,showpacs,preprintnumbers,amsmath,amssymb,10pt]{revtex4-2}

\usepackage{graphicx}% Include figure files
\usepackage{dcolumn}% Align table columns on decimal point
\usepackage{setspace}
\usepackage{bm}% bold math
\usepackage{multirow}
\usepackage{orcidlink}
\begin{document}

\title{Neutron dynamics in the fusion of neutron-rich O ions with $^{12}$C}
\author{A. Leon\,\orcidlink{0009-0001-0462-4203}}
\author{H. Desilets\,\orcidlink{0009-0009-4348-6526}}
\author{Rohit Kumar\,\orcidlink{0000-0002-0450-7218}}
\author{J. Samol\,\orcidlink{0009-0003-4248-3733}}
\author{Z. Berend\,\orcidlink{0009-0002-7871-0654}}
\author{R.~T. deSouza\,
\orcidlink{0000-0001-5835-677X}}
\email{desouza@iu.edu}
\affiliation{%
Department of Chemistry and Center for Exploration of Energy and Matter, Indiana University\\
2401 Milo B. Sampson Lane, Bloomington, Indiana 47408, USA}%

\author{C. Ciampi\, \orcidlink{0000-0002-8142-0052}}
\author{D. Ackermann\, \orcidlink{0000-0001-6284-1516}}
\author{A. Chbihi\, \orcidlink{0000-0001-5653-4325}}
\affiliation{GANIL, CEA/DRF-CNRS/IN2P3, \\
Blvd. Henri Becquerel, F-14076, Caen, France} %

\author{M. Basson\, \orcidlink{0009-0008-8338-6450}}
\author{S. Brown\, \orcidlink{0009-0007-9351-3452}}
\author{K.~W. Brown\, \orcidlink{0000-0003-1923-3595}}
\author{J. Cory\, \orcidlink{0009-0004-8563-0745}}
\author{G. Flores\,}
\author{C.E. McCormick\, \orcidlink{0009-0004-9083-2122}}
\author{V. Zerbach\,\orcidlink{0009-0005-4085-964X}}

\affiliation{Facility for Rare Isotope Beams and Department of Chemistry, Michigan State University, East Lansing, MI 48823, USA} %

\author{C. Dembski\, \orcidlink{0000-0001-9171-9421}}
\author{P.D.O'Malley\, \orcidlink{0000-0001-8793-7653}}
\author{W.W. von Seeger\, \orcidlink{0000-0001-8133-5870}}

\affiliation{Department of Physics and Astronomy, University of Notre Dame, 225 Nieuwland Science Hall, Notre Dame, IN. 46556, USA}

\author{K.Godbey\, \orcidlink{0000-0003-0622-3646}}
\affiliation{
FRIB Laboratory, Michigan State University, East Lansing, Michigan 48824, USA}

\author{A.S. Umar\, \orcidlink{0000-0002-9267-5253}
}
\affiliation{
Department of Physics, Vanderbilt University, Nashville, Tennessee, USA}

\date{\today}
\begin{abstract}
The fusion excitation function for $^{21}$O + $^{12}$C was measured for the first time and compared to the fusion of less neutron-rich isotopes.
The impact of valence neutrons in the d$_{5/2}$ shell on the fusion excitation function is examined. The experimental data manifest a clear dependence of the extracted barrier height, V$_B$, and barrier position, R$_B$ on neutron excess. To assess the role of dynamics the experimental data are compared with both density constrained frozen Hartree Fock (DCFHF) and density constrained time-dependent Hartree Fock (DCTDHF) theories.
\end{abstract}


\maketitle

\textit{Introduction} Nuclei are fascinating two-component quantum drops at the heart of every atom. The field which binds this droplet depends on the ratio of the constituent neutrons and protons. 
One manifestation of the density dependence of the symmetry energy is the existence of a neutron-skin \cite{thiel2019,seif2024}. The magnitude of the neutron skin can thus be used as a probe of the isovector interaction \cite{thiel2019}.
Controlled isotopic comparisons therefore offer a stringent testbed of whether a theoretical approach can consistently describe barrier formation across systems with varying neutron-rich surface properties.
A dynamic probe of this two-component nature is the giant dipole resonance (GDR) \cite{Baldwin47} in which protons and neutrons oscillate out-of-phase \cite{Bortignon98}. 
While the existence of the GDR has been reported for nuclei as light as $^{22}$Ne \cite{Varlamov02}, for such light nuclei, the extraction of the GDR is complicated by the existence of cluster states and pairing effects \cite{He14}. 
Fusion reactions also probe the two-component nature. As two nuclei come together to amalgamate, formation of a low-density neutron-rich neck initially occurs, starting isospin transport.
On qualitative grounds the formation and neutron-richness of this neck should be sensitive to the density dependence of the isovector term in the nuclear equation-of-state \cite{reinhard2016a,arik2026}. Superimposed on this macroscopic binding are the quantum modifications of shell structure, pairing and higher order multi-particle correlations.

Fusion of neutron-rich isotopes of oxygen provides an excellent environment to investigate neutron collectivity. 
The tight binding of the double closed-shell for $^{16}$O resists deformation as the two nuclei approach and fuse. 
Additional neutrons populate the next shell (d$_{5/2}$), and if weakly coupled to the core, might exhibit enhanced neutron dynamics as the fusion proceeds. 
This expectation is supported by considering the shell gap for $N$=8. For example, the difference in the two neutron separation energy at $N$=8 (S$_{2n}$($^{18}$O)-S$_{2n}$($^{16}$O)=
16.7 MeV) is considerably larger than the gap at $N$=20 (
S$_{2n}$($^{42}$Ca)-S$_{2n}$($^{40}$Ca)=
9.1 MeV) and $N$=28 
(S$_{2n}$($^{50}$Ca)-S$_{2n}$($^{48}$Ca)=5.7 MeV). In this work we focus on the impact of d$_{5/2}$ neutrons on fusion of oxygen isotopes with carbon.

\textit{Experimental details} 
In March 2026 the first fusion measurement with a beam of $^{21}$O (t$_{1/2}$= 3.4 s) was made possible by
the Facility for Rare Isotope Beams (FRIB)  at Michigan State University.
A $^{21}$O beam was produced
by bombarding a 10 mm-thick carbon target with a $^{40}$Ar primary beam at E/A = 190 MeV. 
Reaction products were filtered by the ARIS spectrometer before being transported to a linear gas-stopper cell where they were thermalized, extracted, and then charge-bred in the ion trap EBIT.  
Re-acceleration by the ReA3 linac provided a high-quality $^{21}$O beam for experiment E23005  with a sustained intensity of $\sim$800-1000 ions/s at E/A= 3.2 MeV.

The principal element of the setup was the active target detector MuSIC@Indiana, a transverse-field, Frisch-gridded, Multi-Sampling Ionization Chamber \cite{Johnstone21, Desilets25}. 
Segmentation of its anode into twenty 12.5 mm wide strips oriented transverse to the beam direction allowed measurement of the ionization profile produced by the beam or any heavy charged reaction product. 
In this experiment, CH$_4$ gas (99.99\% purity),  served both as a target and the detection medium. 
The gas volume of MuSIC@Indiana was separated from the high-vacuum of the upstream beamline 
by a 2.6 $\mu$m thick mylar window. 
Periodically during the experiment the incident energy of the $^{21}$O, as well as its energy loss in the target gas, were measured by inserting a silicon detector into the active volume from downstream. 
The MuSIC approach is highly efficient as it provides a self-normalized, energy and angle-integrated measurement of the fusion cross-section at multiple energies simultaneously \cite{Carnelli15}. 

In this experiment, MuSIC@Indiana was operated at a pressure 147$\leq$P$\leq$163 torr. 
The charge deposited onto the anode strips by ionizing particles traversing the gas volume was processed by 
conventional charge-sensitive preamplifiers \cite{zepto} and shaping amplifiers before being digitized by Caen V785 peak-sensing ADCs. Digitized signals were readout by a standard PC-based data acquisition system.
All events with incident ions depositing at least 1.0 MeV in the detector were recorded for subsequent analysis. 
A more detailed description of the design, performance, operation, and calibration of MuSIC@Indiana has been previously published \cite{Johnstone21, Johnstone22, Desilets25}. 

While the FRIB experiment examined the most neutron-rich isotope investigated, prior measurements played an important role in characterizing the systematic behavior across the isotopic chain.
Before the FRIB experiment, fusion measurements of $^{19}$O (t$_{1/2}$= 26.5 s) and $^{20}$O (t$_{1/2}$= 13.5 s)  on carbon were conducted at GANIL (Caen, France) using the same experimental setup \cite{Hudan24} extending earlier stable beam measurements of $^{17,18}$O at the University of Notre Dame (UND) \cite{Johnstone21, Johnstone22, Hudan23}. 
Prior to each experiment the fusion excitation function of $^{18}$O + $^{12}$C was measured as a consistency reference. All these reference measurements were found to be in good agreement with both prior
thin target \cite{steinbach2014,Steinbachthesis16,Eyal76,Kovar79} and prior MuSIC@Indiana measurements \cite{Johnstone21, Johnstone22}.
Existing high-quality measurements of the fusion excitation function for $^{16}$O + $^{12}$C reported in the literature were used \cite{Cujec76, Frawley82}.
These systematic measurements of the isotopic chain, carried out at three different laboratories  provide a unique and rich dataset to explore fusion of neutron-rich oxygen isotopes.

Incident ions of $^{21}$O in the FRIB experiment were identified and cleanly separated from beam contaminant $^{21}$F and $^{21}$Ne ions by examining the two-dimensional distribution of $\Delta$E(A1) {\em vs} $\Delta$E(A2), the energy deposit on the initial anodes A1 and A2. 
This selection additionally allowed rejection of reactions occurring from the entrance mylar window. At the near-barrier energies investigated fusion of the $^{21}$O and $^{12}$C nuclei results in a $^{33}$Si compound nucleus (CN) excited to $\sim$40-50 MeV. This CN de-excites to an evaporation residue (ER) by particle emission. Due to the relatively low excitation of the CN, and its excess neutrons, the de-excitation proceeds predominantly by neutron emission. Hence, Z$_{ER}$ is significantly larger than Z$_{BEAM}$, making identification of fusion straightforward. 
Determination of the fusion cross-section requires measuring the number of fusion events relative to the number of incident $^{21}$O ions and accounting for the target thickness. 
The occurrence of fusion is signaled by a sharp increase in the specific ionization relative to the beam, resulting in 
a significant increase in deposited energy, $\Delta$E, on subsequent anodes. 
As the ER slows and stops in the gas the deposited energy peaks and then decreases monotonically \cite{Johnstone21, Desilets25}. 
As any ER produced is measured, no efficiency correction is required in extracting the cross section. 
As in prior analyses, fusion events were cleanly distinguished from two-body scattering events (e.g. inelastic scattering and transfer) as well as proton capture. 
Details describing the characterization of fusion events and the extraction of the fusion cross section have been previously published \cite{Johnstone21,Desilets25}. 

\begin{figure}
\begin{center}
\includegraphics[width=0.45\textwidth]{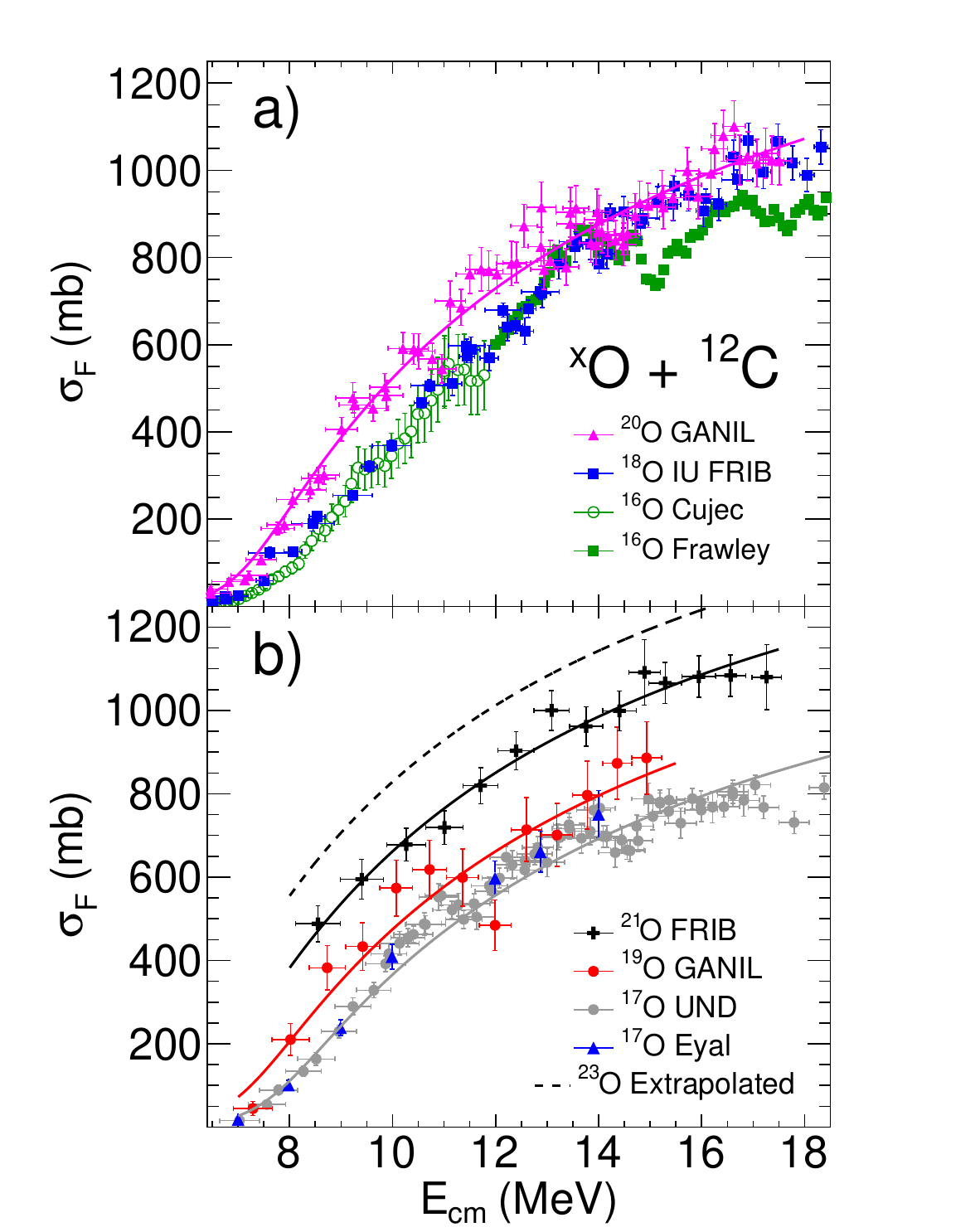}
{}
\caption{Panel a) Fusion excitation functions for oxygen isotopes $^{16,18,20}$O + $^{12}$C. Panel b) Fusion excitation functions for $^{17,19,21}$O + $^{12}$C. The various datasets used are: $^{16}$O \cite{Cujec76} and \cite{Frawley82}; $^{18}$O \cite{Steinbach14a},  \cite{Johnstone21}; $^{20}$O \cite{Desilets25}; $^{17}$O \cite{Hudan23} and \cite{Eyal76}; $^{19}$O \cite{Desilets25a}; $^{21}$O is from FRIB E23005. Solid curves correspond to Wong fits of the experimental data. The dashed curve represents the extrapolated fusion cross section for $^{23}$O based on the neutron-excess dependence of barrier parameters determined in this work.
}
\label{fig:Evens_Odds_with_fit}
\end{center}
\end{figure}

\textit{Results and Discussion} As the presence of an unpaired neutron might impact fusion, the even and odd isotopes are considered separately.
Presented in Fig.~\ref{fig:Evens_Odds_with_fit}a are the fusion excitation functions of $^{16,18,20}$O on $^{12}$C.  
All the excitation functions manifest an overall 
increase with increasing incident energy that is indicative of a barrier-controlled phenomenon.
At essentially all energies presented the $^{20}$O manifests a noticeably larger cross section as compared to $^{16}$O. 
This result might be qualitatively expected based upon the presence of the additional neutrons in $^{20}$O. 

Examination of the $^{16}$O excitation function reveals the presence of several structural features, superimposed on the overall increase in cross section with increasing energy. 
The largest feature for $^{16}$O is a dip at $\sim$15.1 MeV. A dip is also observed for $^{20}$O although it is at a slightly lower energy of $\sim$14.3 MeV, and is smaller in magnitude. Other features also appear to be attenuated for $^{20}$O.
The behavior of the fusion cross section for $^{18}$O is extremely interesting. At low energies its cross section is comparable to that of $^{16}$O while at high energies it is similar to that of $^{20}$O.
This behavior suggests that opposing factors are present for the $^{18}$O excitation function complicating its interpretation.

The structural features evident in fusion of $^{16}$O with $^{12}$C are attributed to the double closed-shell structure of the $^{16}$O nucleus. 
The persistence of these structural features suggests that the strength of the coupling between the valence neutrons in the d$_{5/2}$ shell and the $^{16}$O core is somewhat weak allowing the neutron-rich isotopes to be   
envisaged as a $^{16}$O core  with valence neutrons weakly coupled to it.

The behavior of fusion for incident ions which all have an unpaired neutron is examined in 
Fig.~\ref{fig:Evens_Odds_with_fit}b. 
The fusion excitation functions for $^{17,19,21}$O + $^{12}$C depicts fusion when one, three, and five valence neutrons occupy the d$_{5/2}$ level. 
From $^{17}$O to $^{21}$O the presence of the four additional neutrons provides a significant increase in the cross section at all energies. 
At the highest energies measured the cross section increases from $\sim$800 mb for $^{17}$O to $\sim$1050 mb for $^{21}$O. Neglecting the dip in the cross section at $\sim$12 MeV for $^{19}$O  its cross section is intermediate between that of $^{17}$O and $^{21}$O as might be expected.  
The high quality of the $^{17}$O MuSIC data \cite{Hudan23}, extended with more recent measurements, is reinforced by their good agreement with the thin target data depicted by the triangles \cite{Eyal76}.

For even neutron number oxygen nuclei, extraction of the impact of neutron excess on fusion is complicated by pairing and the Pauli exclusion principle which act to suppress fusion \cite{Simenel17, Tong23}. For the remainder of this work we therefore focus on fusion for isotopes with an unpaired neutron namely $^{17,19,21}$O. We also consider $^{20}$O as the most neutron-rich nucleus previously measured.

To compare fusion for different systems various scalings have been developed \cite{Canto15, Jiang24}. 
In the universal fusion function the $\ell$-dependent potential between two heavy-ions as they fuse is accounted for
through an effective partial wave. 
Although a description of fusion by a single universal function is intriguing, in this work a simpler approach is adopted. 
This approach is justified as only a single isotopic chain at near barrier energies is examined. 
In order to extract the systematic behavior of the fusion cross section with the number of valence neutrons, the measured fusion excitation functions are fit with the Wong fusion formula \cite{Wong73}:

\begin{equation}
\sigma_W = \frac{\hbar\omega R_B^2}{2E}ln\{1+ exp(\frac{2\pi}{\hbar\omega}(E-V_B)\}
\end{equation}

The fusion cross section, $\sigma_W$, is controlled by the presence of a parabolic barrier
with height, V$_B$, position R$_B$, and width $\hbar\omega$.
Limited by the low intensity of radioactive beams, the present work is restricted to near/above barrier energies and thus is largely insensitive to the barrier penetrability, dictated by $\hbar\omega$. 
The measured cross section is thus principally sensitive to the height, V$_B$  and the position, R$_B$ of the barrier. 
Well above the barrier the cross section is mainly determined by R$_B$ while near the barrier both V$_B$ and R$_B$ play a role in determining the cross section.  
A positive correlation exists between these two quantities. 
A larger
value for V$_B$ reduces the flux overcoming the barrier. 
To obtain the same cross section, an increase in V$_B$ must be balanced by a larger value of R$_B$, corresponding to a more extended configuration at the saddle point. 

While this simple description of fusion by a single barrier ignores the presence of multiple angular momentum 
dependent barriers \cite{Esbensen12, Simenel13}, it can be justified in the present work by the investigation being limited to comparison of nuclei in an isotopic chain and a relatively narrow energy range where only a small number of incident $\ell$-waves ($\ell$$\leq$20) result in fusion \cite{deSouza24}. Limited low-energy data in the case of $^{19}$O and $^{21}$O led to fixing the value of $\hbar\omega$ for the odd isotopes to 3.0 MeV. Varying $\hbar\omega$ within reasonable limits, 2.0$\leq$$\hbar\omega$$\leq$4.0, does not significantly affect the cross section in the measured regime.
The Wong fits for the different isotopes are presented in Fig.\ref{fig:Evens_Odds_with_fit} as solid lines, with the values of the fit parameters indicated in Table \ref{tab:Wong_param}. 
In all cases reasonably good fits were achieved.

\begin{table}
    \centering
    \begin{tabular}{ccccccc}
        Nuclide & & V$_B$(MeV) & & R$_B$(fm) & & $\hbar\omega$(MeV)\\
        \hline\\
%         $^{16}$O & & 7.58 & & 7.10 & & 2.25\\
         $^{17}$O & & 7.58 & & 6.93 & & 3.00 \\
%        $^{18}$O & & 7.86 & & 7.70 & & 3.18\\
         $^{19}$O & & 7.03 & & 7.13 & & 3.00\\
         $^{20}$O & & 7.02 & & 7.48 & & 2.74\\
         $^{21}$O & & 6.31 & & 7.56 & & 3.00\\
    \end{tabular}
    \caption{Values of fit parameters resulting from Wong fits of the experimental data.}
    \label{tab:Wong_param}
\end{table}

\begin{figure}
\begin{center}
\includegraphics[width=0.50\textwidth]{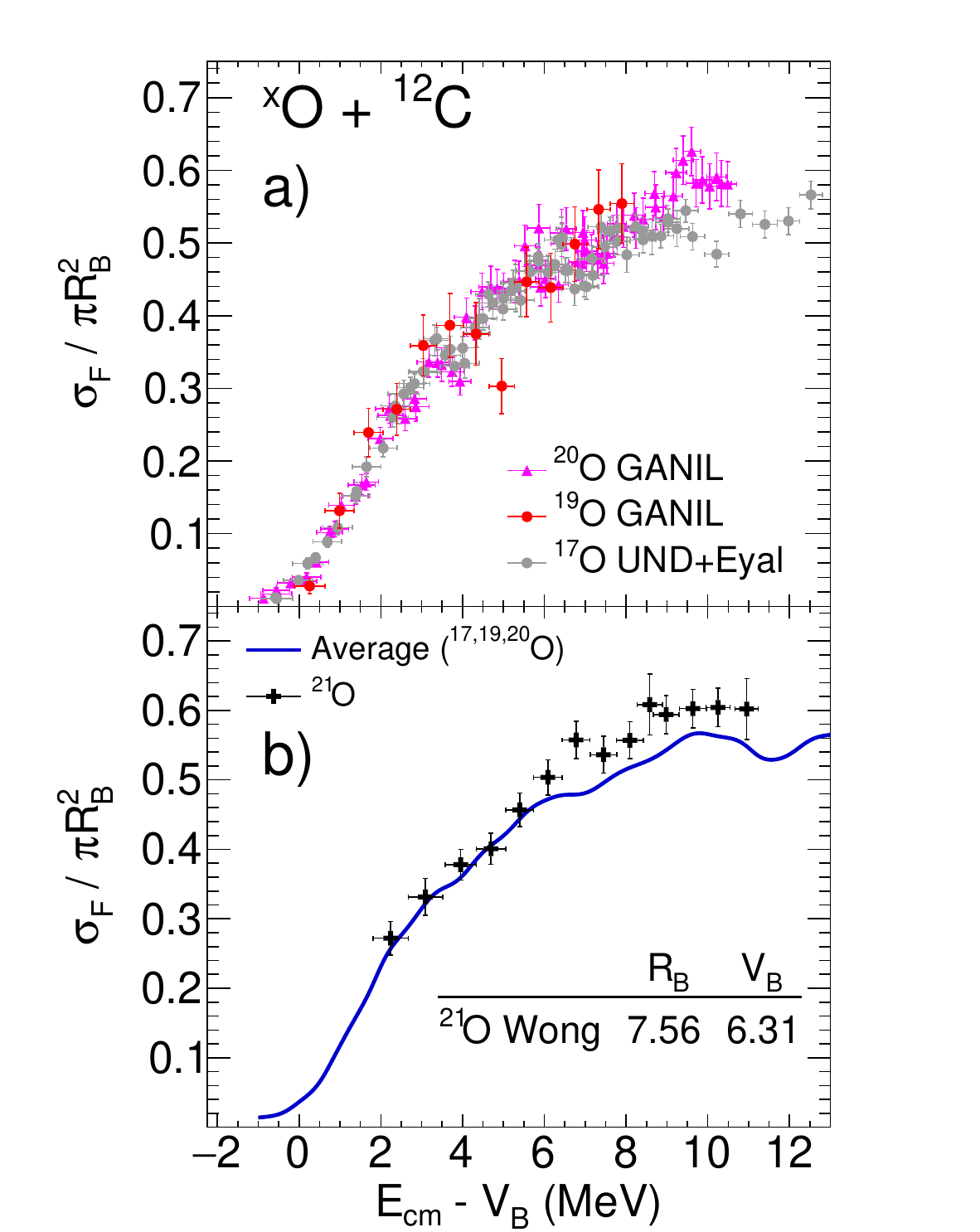}
\caption{Top: Dependence of the reduced fusion excitation function for $^{17}$O, $^{19}$O, and $^{20}$O ions on $^{12}$C on above-barrier energy. Bottom: Comparison of the average reduced fusion excitation function depicted in the top panel with the reduced fusion excitation function for $^{21}$O. }
\label{fig:Reduced_Xsect}
\end{center}
\end{figure}

To assess the effectiveness of extracting V$_B$ and R$_B$ the dependence of the reduced cross section, $\sigma_F$/$\pi$R$_B^2$, on above barrier energy (E$_{cm}$-V$_B$), is examined in Fig.~\ref{fig:Reduced_Xsect}.  
In Fig.~\ref{fig:Reduced_Xsect}a isotopes $^{17,19,20}$O are presented. 
One observes that the cross section curves are in reasonably good agreement. 
This collapse onto a common curve suggests that the extracted V$_B$ and R$_B$ values are sound. 

In Fig.~\ref{fig:Reduced_Xsect}b the reduced excitation function for $^{21}$O (solid symbols) is compared with the average reduced excitation derived from $^{17,19,20}$O. 
To facilitate comparison the average reduced excitation function obtained in 
Fig.~\ref{fig:Reduced_Xsect}a was smoothed by a
gaussian ($\sigma$=0.35) and  is indicated by the solid line. 
The Wong fit for $^{21}$O results in V$_B$=6.31 MeV and R$_B$=7.56 fm. 
It is apparent from Fig.~\ref{fig:Reduced_Xsect} that the 
$^{21}$O reduced excitation function differs only slightly from that of the systematics obtained for the other members of the isotopic chain. Primarily, $^{21}$O exhibits a slightly smaller R$_B$ than would be expected from the systematic behavior derived.

\begin{figure}
\begin{center}
\includegraphics[width=0.45\textwidth]{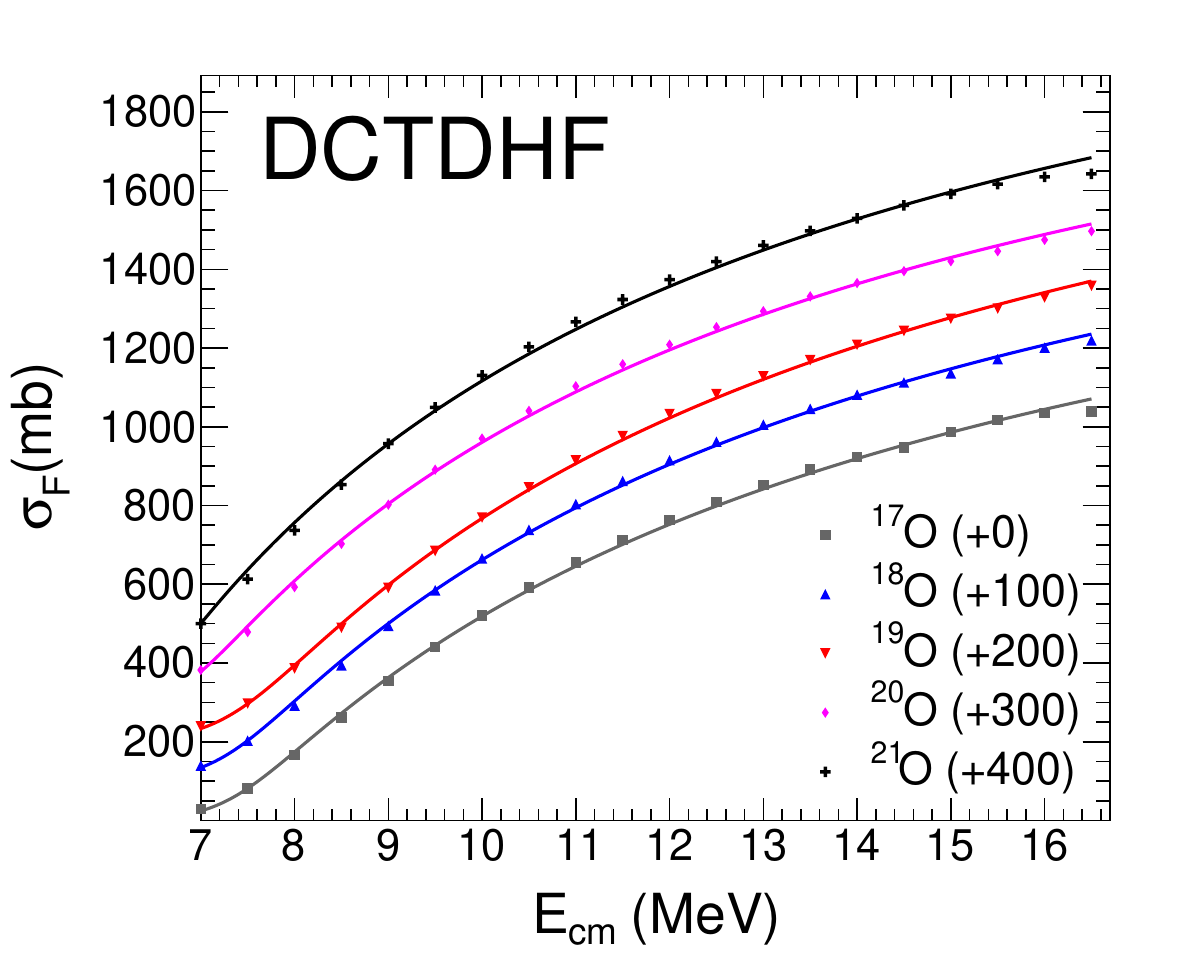}
\caption{Theoretical fusion excitation functions for $^{17-21}$O on $^{12}$C from DCTDHF calculations are indicated as closed symbols. For clarity the excitation function for different isotopes is offset vertically by \mbox{$(A - 17) \times 100$} mb. Solid lines indicate fits of the theoretical results with the Wong formula.}
\label{fig:DCFHF_DCTDHF_theory}
\end{center}
\end{figure}

To examine the role of dynamics introduced by the valence neutrons in the neutron-rich oxygen isotopes we have performed density-constrained frozen Hartree-Fock (DCFHF) and density constrained time-dependent Hartree Fock (DCTDHF) calculations~\cite{simenel2018,simenel2025}. As the density distributions are frozen in the former case they provide the key reference for the impact of mean-field dynamics as realized in the DCTDHF calculations.

TDHF is a many-body approach that is well suited to describe the large-amplitude collective motion associated with fusion while also describing the transfer dynamics, equilibration processes, and Pauli blocking that affect heavy-ion fusion probabilities~\cite{godbey2017,simenel2020,simenel2017}.
These effects are included self-consistently and no fits are performed beyond the original calibration of the energy density functional (EDF). We emphasize that the inter-ion potential is obtained microscopically in TDHF based calculations~\cite{umar2006b,simenel2018}.

The excitation functions predicted by the DCTDHF theory are presented in Fig.~\ref{fig:DCFHF_DCTDHF_theory} as symbols. As expected, the predicted cross sections decrease smoothly with decreasing energy. While the DCTDHF theory provides direct access to the barrier and hence the quantities R$_B$ and V$_B$, it is also instructive to see the extent to which extracting these quantities from the predicted cross sections impacts their values. To accomplish this,  the theoretical predictions were fit with the Wong formula and the resulting fits are displayed in Fig.~\ref{fig:DCFHF_DCTDHF_theory} as the solid lines. A good description of the predicted cross sections is achieved with a Wong fit.

\begin{figure}
\begin{center}
\includegraphics[width=0.45\textwidth]{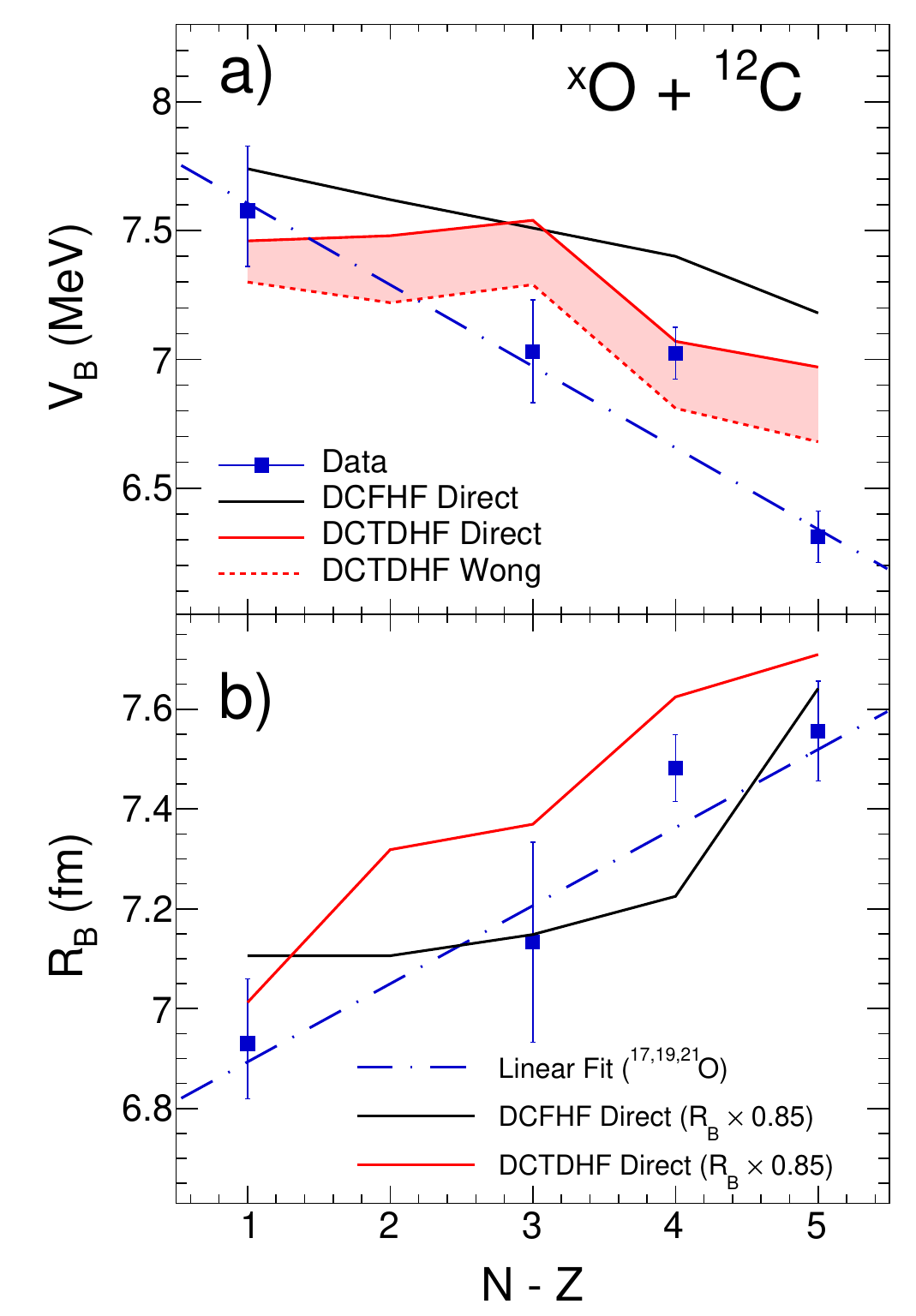}
\caption{Dependence of the fusion barrier height, V$_B$, (top) and radius, R$_B$ (bottom) on neutron excess for fusion of oxygen ions with $^{12}$C. Solid black and red lines correspond to DCFHF and DCTDHF calculations respectively (see text for details).}
\label{fig:Vb_Rb_data_theory}
\end{center}
\end{figure}

In  Fig.~\ref{fig:Vb_Rb_data_theory} the dependence of the extracted barrier parameters, V$_B$ and R$_B$, on neutron excess for the experimental data is presented as symbols. The data show an overall decrease of V$_B$  
with increasing neutron excess.
For nuclides with an unpaired neutron an essentially linear decrease in  V$_B$ is observed. Fitting the data reveals a decrease in V$_B$ of approximately 640 keV/neutron pair as indicated by the dot-dash line.
The extracted barrier for $^{20}$O, lies slightly higher than this line, perhaps as a consequence of pairing.   
 The dependence of R$_B$ on neutron excess is presented in Fig.~\ref{fig:Vb_Rb_data_theory}b.  R$_B$ increases approximately linearly from 6.9 fm (A=17) to almost 7.6 fm (A=21) with a change of four neutrons.  
 
 The error bars shown for V$_B$ and R$_B$ in Fig.~\ref{fig:Vb_Rb_data_theory} represent the uncertainty associated with 2.0$\leq\hbar\omega\leq$4.0. The uncertainty for $^{20}$O is smaller than the other isotopes as the experimental data is measured to lower energies. A larger error bar is assessed for $^{19}$O due to the presence of the resonance at (E$_{cm}$-V$_B$)$\approx$5 MeV.

The extracted Wong barrier parameters from both the DCFHF and DCTDHF theoretical calculations are also presented in Fig.~\ref{fig:Vb_Rb_data_theory}. The solid black lines represent the DCFHF predictions that account for the change in static size of the different isotopes but do not include dynamics. 
These calculations clearly underpredict the decrease of V$_B$ that is experimentally observed.
The prediction of V$_B$ by the DCTDHF theory is indicated by the solid red line. Although the DCTDHF theory, which includes dynamics at the mean field level, shows differences for individual isotopes in V$_B$, essentially the same average slope with neutron excess as the frozen calculations is observed. The DCTDHF calculations do not exhibit a dependence for V$_B$ on neutron excess as strong as the one observed for the experimental data. The impact of using Wong fits to extract V$_B$ from the theoretically predicted cross section is shown as a dotted line. Although the Wong fits systematically yield lower values of V$_B$, the dependence of V$_B$ on neutron excess is largely unchanged.
The value of measuring the extremely neutron-rich nucleus $^{21}$O is clearly apparent as it provides a sensitive test of the dynamics. The DCFHF calculations predict a barrier V$_B$ that is $\sim$0.7 MeV larger for $^{21}$O than is experimentally measured. Although the inclusion of dynamics lessens this difference it still is not in good agreement suggesting that neutron dynamics play a larger role near the barrier than predicted by the DCTDHF theory.

As both DCFHF and DCTDHF overpredict the magnitude of R$_B$ as compared to the experimental data we have scaled the predicted values by 0.85 to facilitate comparison. The DCFHF calculations predict only a very modest increase of $\sim$0.1 fm from $^{17}$O to $^{20}$O followed by an increase of 0.4 fm to $^{21}$O. Inclusion of dynamics leads to a more systematic increase across the isotopic chain with a change of 0.6 fm from A=17 to A=21, roughly comparable to the experimental data.

The systematic behavior with neutron excess in the experimental data allows us to extrapolate the near-barrier fusion cross section for the reaction $^{23}$O + $^{12}$C. As the last bound oxygen isotope with an unpaired neutron we anticipate it to have the largest fusion cross section of the isotopic chain. Using the linear trend for the V$_B$ and R$_B$ in Fig.~\ref{fig:Vb_Rb_data_theory} indicated by the dot-dashed line, we extrapolate that V$_B$= 5.70 MeV and R$_B$= 7.84 fm for $^{23}$O. The Wong excitation function corresponding to these values along with $\hbar\omega$=3 is shown in Fig.~\ref{fig:Evens_Odds_with_fit} as a dashed line. As $^{23}$O has t$_{1/2}$= 96 ms, direct measurement of its fusion excitation function is not likely for the foreseeable future. 

\textit{Conclusions} The fusion excitation function for $^{21}$O + $^{12}$C was measured for the first time and compared to the fusion excitation function for less neutron-rich oxygen isotopes. Parameterization of the measured cross sections with the Wong fusion formula allowed examination of 
the dependence of the extracted barrier height, V$_B$, and barrier position R$_B$ on neutron excess.
For odd-A isotopes  the experimental data barriers manifested a clear decrease with increasing neutron excess. 
For each neutron pair the barrier is lowered by approximately $\sim$ 640 keV. 
A slightly higher barrier than that predicted by the odd-A barrier systematics is observed for $^{20}$O and is attributed to pairing. 
Comparison with theoretical calculations clearly showed that while dynamics at the mean field level, as realized in the DCTDHF calculations, partly explained this decrease, they fail in completely describing it. 
The experimental data also exhibited a clear dependence of the barrier position on neutron excess. 
Both theories significantly over-predicted the position of the barrier. 
If the predicted barrier position is scaled, inclusion of dynamics does correctly predict the change in the barrier position with neutron excess. 
These results indicate that  
fusion occurs at smaller internuclear separation than predicted by the theory and that it is more sensitive to dynamics than predicted by the DCTDHF theory with the largest difference observed for $^{21}$O.
This work demonstrates that systematic characterization of the near barrier behavior of the excitation function for neutron-rich nuclei can be a powerful tool in understanding neutron dynamics  and motivates the measurement of 
fusion for other neutron-rich nuclei, in particular $^{22}$O.

%%\begin{table} [t!]

%%  \begin{center}
%%    \begin{tabular}{ccccccc}
%%    \hline
%%    &$^{16}$O & $^{17}$O & $^{18}$O & $^{19}$O &  $^{20}$O & $^{21}$O\\
%%    \hline
%%    1 n &-10717& 803 & -3099 & 990 & -2661 & 1141 \\
%%    2 n &-15765& -6684 & 934 & 1122 & 1559 & 1709 \\
%%    \end{tabular}
%%  \end{center}
%%  \caption{Q-values (in keV) for single neutron transfer from different oxygen isotopes to $^{12}$C.}
  
%%  \label{table:ntrans} 
%%\end{table}

\textit{Acknowledgements} We acknowledge the high-quality beam and experimental support provided by the technical and scientific staff at the Facility for Rare Isotope Beams (FRIB), in particular the ReA3 staff.
We are thankful for the high-quality services of the Mechanical Instrument Services and Electronic Instrument Services facilities at Indiana University.
This work was supported by the U.S. Department of Energy Office of Science under Grant No. 
DE-SC0025230 (Indiana University), DE-SC0021938 (MSU), DE-SC0022299 (MSU), DE-SC0013847 (Vanderbilt University), DE-SC0023175 (Office of Science, NUCLEI SciDAC-5 collaboration), DOE-DE-NA0004074 (NNSA, the Stewardship Science Academic Alliances program) and the National Science Foundation under Grant Nos. PHY-2310059 (University of Notre Dame) and PHY-2309923 (Michigan State). 
C.E.M. acknowledges support from the National Science Foundation Graduate Research Fellowship (GRF) under Grant No. 2235783 and DGE-2236418.
This material is based upon work supported by the U.S. Department of Energy, Office of Science, Office of Nuclear Physics and used resources of the Facility for Rare Isotope Beams (FRIB) Operations which is a DOE Office of Science User Facility under Award Number DE-SC0023633.

\textit{Data Availability} -
The data that support the findings of this article are not publicly available upon publication because it is not technically feasible and/or the cost of preparing, depositing, and hosting the data would be prohibitive within the terms of this research project. The data are available from the authors upon reasonable request.

%\bibliography{O21C12_FRIB,VU_bibtex_master}

\end{document}